\documentclass{jltp}

\usepackage{graphicx} 
\def\be{\begin{equation}}
\def\ee{\end{equation}}

\title{Spin-Current Relaxation Time in Spin-Polarized Heisenberg Paramagnets}  

\author{R. Ragan, K. Grunwald, and B. Batell}

\address{Department of Physics\\
University of Wisconsin at La Crosse, La Crosse WI 54601}

\runninghead{R. Ragan, K. Grunwald, and B. Batell}{Spin-Wave Relaxation Time in Spin-Polarized Heisenberg Paramagnets}

\begin{document}
\maketitle
\begin{abstract}
We study the spatial Fourier transform of the spin correlation function $G_q(t)$ in paramagnetic quantum crystals by direct simulation of a 1d lattice of atoms interacting via a nearest-neighbor Heisenberg exchange Hamiltonian. Since it is not practical to diagonalize the $s=1/2$ exchange Hamiltonian for a lattice which is of sufficient size to study long-wavelength (hydrodynamic) fluctuations, we instead study the $s \rightarrow \infty$ limit and treat each spin as a vector with a classical equation of motion.  The simulations give a detailed picture of the correlation function $G_q(t)$ and its time derivatives. At high polarization, there seems to be a hierarchy of frequency scales: the local exchange frequency, a wavelength-independent relaxation rate $1/\tau$ that vanishes at large polarization $P \rightarrow 1$, and a wavelength-dependent spin-wave frequency $\propto q^2$. This suggests a form for the correlation function which modifies the spin diffusion coefficients obtained in a moments calculation by Cowan and Mullin, who used a standard Gaussian ansatz for the second derivative of the correlation function.
\end{abstract}

The Heisenberg spin chain continues to be of the focus of much experimental and theoretical study (See Ref.\onlinecite{Stark} and references therein). A number of new materials has been synthesized which appear to be nearly perfect experimental realizations of the spin-1/2 chain.  These experiments have also pointed to the need for detailed theroeticals studies of the spin dynamics of the Heisenberg chain. In its simplest form the model is defined by the Hamiltonian
\be
H=\sum_{i,j}J{\bf S}_{i} \cdot {\bf S}_{j}-\sum_{i} BS^{z}_i
\label{Ham}
\ee
where $B$ is the external magnetic field, $J$ is the nearest neighbor exchange integral, the second sum is over nearest neighbors. Its low temperature ($T\ll J$) behavior is controlled by the $T=0$ quantum critical point and is well-described by quantum field theory. The finite temperature dynamics are less understood. The infinite temperature regime has been studied theoretically, and numerically, but almost exclusively in the unpolarized $B=0$ case. In this paper, we focus on highly polaralized paramagnet at infinite temperature $J/T \rightarrow 0$, but with $B/T$ finite.  The 3-d version of the spin-1/2  system has been proposed as a model of spin transport in the paramagnetic hcp phase of solid $^3$He.\cite{Cowan94}$^-$\cite{Cowanun}

Previously, we studied spin transport\cite{Ragan} in a 2-d lattice model that demonstrated  Leggett-Rice type behavior, i.e., characterized by an exchange mean-field supporting damped spin-waves. The Leggett equation\cite{Legg70} was originally derived for paramagnetic Fermi fluids, but it has been shown that it applies to the spin dynamics of any polarized system interacting via quantum exchange.\cite{Meyer90} Whereas in the liquid longitudinal spin transport is due to particle motion and damping of the spin current is due to particle collisions, in a lattice of localized spins transport is due to exchange and damping is due to spin fluctuations.  In the current paper we study 1-d chains and find another analogy with the fluid: a relaxation of the spin-current at high polarization, where the spin system begins to acts like a dilute gas with a mean free time $\tau$. We restrict our attention to longitudinal spin diffusion, although a similar description seems to apply for transverse spin transport (the results are of a more preliminary nature, however). Specifically, we study the polarization dependence of spatial Fourier transform of the spin correlation function $G_q(t)$ in the hydrodynamic limit $q \rightarrow 0$. The time derivatives of this function (or the moments of its Laplace transform) give the diffusion coefficient $D_{\parallel}$, and the relaxation time $\tau$.  At high polarization, there seems to be a hierarchy of frequency scales: the local exchange frequency $J/\hbar$, a wavelength-independent relaxation rate $1/\tau(P)$ that vanishes at large polarization $P \rightarrow 1$, and a diffusive decay rate $\gamma=D_{\parallel} q^2$. 

Since a lattice of $N$ spins involves $2^{N}$ quantum states, it is not practical to diagonalize the $S={1\over 2}$ exchange Hamiltonian for a lattice which is of sufficient size to study long-wavelength (hydrodynamic) fluctuations. Instead, We take a commonly used approach \cite{Wang}$^{-}$\cite{quantum} and study the $S \rightarrow \infty$ limit and treat each spin as a vector ${\bf S}({\bf r})$ with a classical equation of motion, 
\be
\frac {d}{dt}{\bf S}({\bf r})=\Omega \sum_{{\bf r}^{\prime}} {\bf S}({\bf r}) \times {\bf S}({\bf r}^{\prime})
\label{motion}
\ee
where the sum is over nearest neighbors and  $\Omega=J/\hbar$ is the exchange frequency. For convenience, we have moved to the Larmor frame, and have set the vector magnitude of the spins $|{\bf S}|$ equal to $1$. Also, we found only quantitative differences between 1 and 2-d systems, so we have concentrated on 1-d systems, which are computationally less expensive, and for which theoretical results are easier to obtain. Although it is known\cite{Love} that the unpolarized chain exhibits anomolous diffusion $\gamma_q\propto q^{\alpha}$ with $\alpha$ slightly less than 2, this behavior does not seem to be a present at higher polarizations, where the fluctuations essentially pass through one another. 

Longitudinal diffusion was studied by simulating the motion of a hydrodynamic ($q \rightarrow 0$) fluctuation $S^{z}_{q}(t)$. We define the correlation function\cite{Forster}
\be 
G_{q}(t)=\frac{\langle S^{z}_{q}(t) S^{z*}_{q}(0)\rangle }{\langle S^{z}_{q}(0) S^{z*}_{q}(0)\rangle }
\label{corr}
\ee
which is assumed to have an exponential decay $\propto e^{-D_{\parallel}q^{2}t}$ as $t\rightarrow \infty$. The brackets indicate an average over initial conditions sampled from an equilibrium ensemble.
\begin{figure}
\includegraphics[width=5.0in]{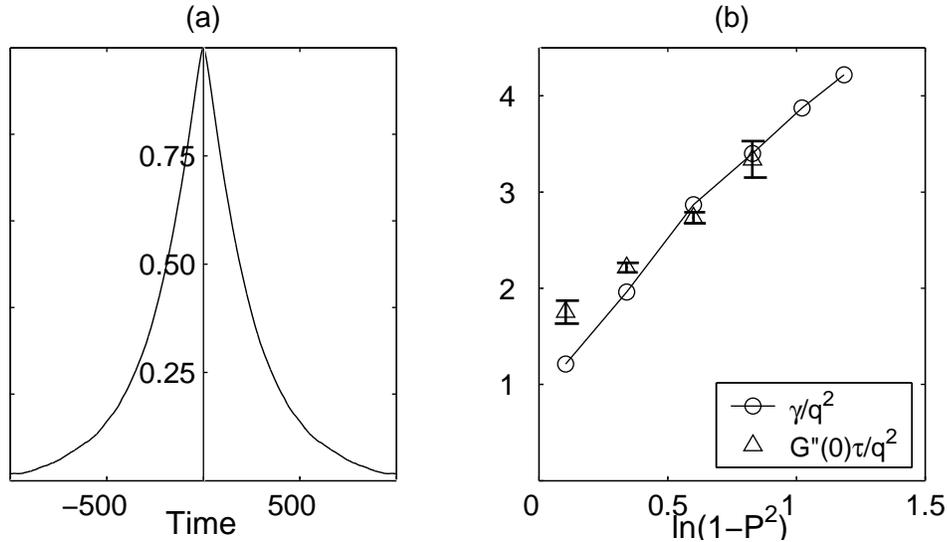}
\caption{(a) Spin correlation function for $P=53.7\%$. (b) Log-Log plot of the spin diffusion coefficient $D_{\parallel}$ vs. $1-P^2$. form the longtime diffusive behavior and the microscopic relaxation time of $G_q''(t)$. The last four points have a slope of $2.3\pm 0.1$}
\label{fig1}
\end{figure}

We briefly outline the procedure for calculating $G_{q}(t)$ and its derivatives. First, an initial spin configuration was obtained by generating a lattice of randomly oriented spins ${\bf S}_{i}$ with a Boltzman distribution $e^{(B/T)S^{z}_{i}}$. The exchange energy was neglected in the Boltzman factor, so the spins were initially uncorrelated.  The spin configuration was then allowed to evolve according to Eq.\ref{motion}, and the spin correlation function $G_{q}(t)$ was calculated
from the definition Eq.\ref{corr}. The average was then performed over $N$ initial configurations where $N=10^6$ was typical.  Periodic boundary conditions were used and the length $L$ of the lattice was chosen to be much greater than the \lq\lq mean free path" given by $c \tau$ where $c=2\Omega a$ is the maximum spin wave phase velocity, ($a$ is the lattice spacing) and $\tau$ is the "mean free time" (defined below). The equation of motion was integrated with a fourth-order Runge-Kutta method, with time step $\Delta t = 0.1$. The typical motion of a spin in a time step was 5 degrees or so. The conserved quantities such as the total energy, the polarization, and the individual spin magnitudes were found to be constant to a high degree of accuracy. 

\begin{figure}
\includegraphics[width=5.0in]{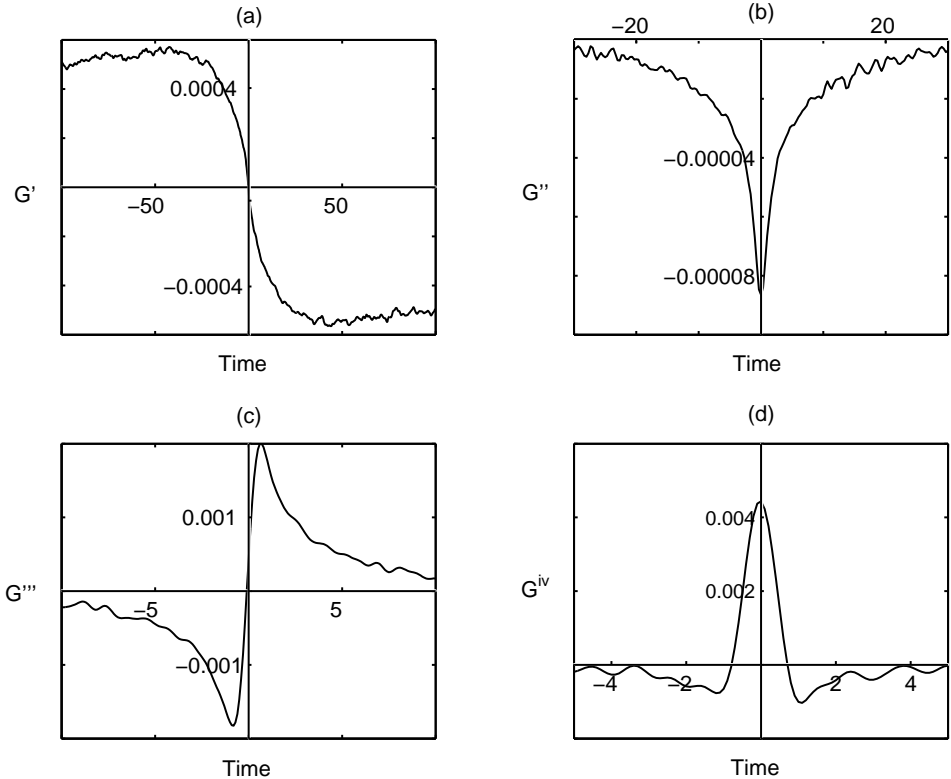}
\caption{First four derivatives of the correlation function shown in Fig.1a. Note the hiearchy of time scales.}
\label{fig2}
\end{figure}
 In simulations to study the short time behavior of $G_{q}(t)$ and its derivatives the system was allowed to relax for a time greater than the microscopic relaxation time $\tau$ to remove any artifical effects of the uncorrelated initial conditions. The higher derivatives were, of course, a good deal noisier than $G_q(t)$ data and up to $N=10^8$ simulations were even averaged in some cases, although these were short simulations to study short-time behavior.

In Fig.1a the correlation function $G^{z}_{q}(t)$ is shown for $P=53.7\%$ and is indistinguishable from $\exp(-\gamma_q|t|)$. From this function we extract the spin diffusion coefficient $D_{\parallel}=\gamma_q/q^2.$ The polarization dependence is shown in Fig.1b, from which we find $D_{\parallel} \propto (1-P^2)^{-2.3\pm0.1}$. We also get a detailed picture of the short-time behavior of the correlation function. In Figs.2a-d we show the first four derivatives of $G_q(t)$. The surprising result is the fact that at high polarization the second derivative is not Gaussian but has the form of a hydrodynamic correlation function $G_q''(t)=G_q''(0)\exp(-|t|/\tau)$, where $\tau$ is a polarization dependent relaxation time that diverges as $P\rightarrow 1$ and $G_q''(0)\propto q^2$. (It is easy to show from Eqs.(2-3) that $G_q''(0) \rightarrow -2\Omega^2 a^2 q^2$ as $P\rightarrow 1$.) Accordingly, the first derivative relaxes {\it exponentially} to its quasi-steady value of $-\gamma_q G_q(t_{c1})$, where $t_{c1}$ is a cut off time $\tau\ll t_{c1}\ll \gamma_q^{-1})$. Thus, we have the relation
\be
G'_q(t_{c1})=-D_{\parallel}q^2G_q(t_{c1})=\int_0^{t_{c1}}G''_q(t)dt=G''(0)\tau
\label{G2tau}
\ee
giving $D_{\parallel}=G''(0)\tau/q^2$ which relates the hydrodynamic diffusive behavior to the  microscopic behavior.

The fourth derivative evolves on an even shorter (in fact the shortest possible) time scale $\sim 1/\Omega$. For large polarizations $G_q^{iv}(t)$ has the form  form $G_q^{iv}(t)=G_q^{iv}(0)f(P,t)$, where $G_q^{iv}(0)\propto q^2$ is polarization dependent and $\rightarrow 0$ as $P \rightarrow 1$. The function $f(P,t)->0$ (more precisely to a value $\sim G_q''(0)/\tau^2 \ll G_q^{iv}(0)$) on a time scale $t_{c2}$, where here the cutoff time satisfies $\Omega^{-1}\ll t_{c2} \ll \tau$ (See Fig.2b). Assuming this form for $G_q^{iv}(t)$ we have the relation 
\be
G'''_q(t_{c2})=-\frac{G''_q(t_{c2})}{\tau}=\int_0^{t_{c2}}G^{iv}_q(t)dt=G^{iv}_q(0)F(P)
\label{G4tau}
\ee
Putting this into the expression for $D_\parallel$ we get our main result
\be
D_\parallel \approx \frac{[G_q''(0)]^2}{q^2G_q^{iv}(0)F(P)}
\label{D}
\ee

Next we compare our numerical results to the moments method calculations of Cowan and Mullin {\it et al.}\cite{Cowan94}$^{-}$\cite{Cowanun}$^,$\cite{Cowan89}, who considered a nearest-neighbor pair exchange Hamiltonian to describe the quantum spin-1/2 dynamics of paramagnetic $^{3}$He in its hcp phase, with a nearest-neighbor pair exchange Hamiltonian. Using a Gaussian ansatz for the second derivative they found, for arbitrary polarization, $D_\parallel \propto [G_q''(0)]^{3/2}/[G_q^{iv}(0)]^{1/2}$, which diverges like $(1-P^2)^{-1/2}$ as $P \rightarrow 1$. If we use Eq.(\ref{D}) and their result that $G_q^{iv}(0)\propto (1-P^2)$ as $P \rightarrow 1$, the divergence of $D_{\parallel}$ would be modified, diverging instead at least like $(1-P^2)^{-1}$. This is still much slower than divergence seen in our numerical simulations. However, the divergence will be faster if the integral $F(P) \rightarrow 0$ as $P \rightarrow 1$, which from Fig.2d seems entirely possible.  In any event, our results shift the focus to the calculation of $\tau$ which depends on the short-time ($\Omega^{-1}$) behavior of the fourth derivative.

\section*{ACKNOWLEDGMENTS}
The authors thank W. Mullin and B. Cowan for useful discussions. This research was supported by NSF grant DMR-0209606.


\begin{thebibliography}{99}
\bibitem{quantum} We do not present the details here, but we compared the classical and quantum correlation functions for small paramagnetic systems ($N \le 10$), and the similarity of the two is especially evident in the frequency spectra as $N$ is increased.
\end{thebibliography}
\end{document}